\documentclass[reprint,showpacs,preprintnumbers,amsmath,amssymb,nofootinbib,groupedaddress,superscriptaddress,showkeys,twocolumn]{revtex4-1}

\usepackage{dcolumn}
\usepackage{bm}
\usepackage{fancybox}
\usepackage{mathrsfs}
\usepackage{booktabs}
\usepackage[pdftex]{graphicx}
\usepackage{epstopdf}
\usepackage[bookmarks=true,bookmarksnumbered=true,bookmarkstype=toc]{hyperref}

\newcommand{\lsim}{\raise0.3ex\hbox{$\;<$\kern-0.75em\raise-1.1ex\hbox{$\sim\;$}}}
\newcommand{\gsim}{\raise0.3ex\hbox{$\;>$\kern-0.75em\raise-1.1ex\hbox{$\sim\;$}}}

\allowdisplaybreaks[4]

\newcommand{\lmlt}{L_\mu^{} - L_\tau^{}}

\def\not#1{\ /\hspace{-2.5mm}{#1}}

\begin{document}

\title{
Detecting the $\lmlt$ gauge boson at  
Belle II
}

\author{Takeshi Araki}
\email{araki@krishna.th.phy.saitama-u.ac.jp}
\affiliation{%
Department of Physics, Saitama University,
\\
Shimo-Okubo 255, 
338-8570 Saitama Sakura-ku, 
Japan
}

\author{Shihori Hoshino}
\email{hoshino@krishna.th.phy.saitama-u.ac.jp}
\affiliation{%
Department of Physics, Saitama University,
\\
Shimo-Okubo 255, 
338-8570 Saitama Sakura-ku, 
Japan
}

\author{Toshihiko Ota}
\email{tota@yachaytech.edu.ec}
\affiliation{%
Department of Physics, Saitama University,
\\
Shimo-Okubo 255, 
338-8570 Saitama Sakura-ku, 
Japan
}
\affiliation{%
Department of Physics, Yachay Tech,
\\
Hacienda San Jos\'{e} s/n y Proyecto Yachay, 
100119 San Miguel de Urcuqu\'{i}, 
Ecuador
}

\author{Joe Sato}
\email{joe@phy.saitama-u.ac.jp}
\affiliation{%
Department of Physics, Saitama University,
\\
Shimo-Okubo 255, 
338-8570 Saitama Sakura-ku, 
Japan
}

\author{Takashi Shimomura}
\email{shimomura@cc.miyazaki-u.ac.jp}
\affiliation{%
Faculty of Education, 
University of Miyazaki,
\\
1-1 Gakuen-Kibanadai-Nishi,
889-2192 Miyazaki,
Japan
}

\date{\today}


\preprint{\bf STUPP-17-229, UME-PP-006, YACHAY-PUB-17-01-PN}


 \begin{abstract}
  We discuss the feasibility of detecting the gauge boson of
  the $U(1)_{\lmlt}^{}$ symmetry, which possesses a mass
  in the range between MeV and GeV, 
  at the Belle-II experiment.
  The kinetic mixing between the new gauge boson $Z'$
  and photon is forbidden at the tree level
  and is radiatively induced.
  The leptonic force mediated by such a light boson
  is motivated by the discrepancy in muon anomalous magnetic moment
  and also the gap in the energy spectrum of cosmic neutrino.
  Defining the process $e^{+} e^{-} \rightarrow \gamma Z'
  \rightarrow \gamma \nu \bar{\nu}~(missing~energy)$
  to be the signal,
  we estimate the numbers of the signal and the background events
  and show the parameter region to which the Belle-II experiment
  will be sensitive.
  The signal process in the $\lmlt$ model is enhanced
  with a light $Z'$, which is
  a characteristic feature differing from
  the dark photon models with a constant kinetic mixing.
  We find that
  the Belle-II experiment with the design luminosity
  will be sensitive to the $Z'$ with
  the mass $M_{Z'} \lesssim 1 $ GeV and
  the new gauge coupling $g_{Z'} \gtrsim 8\cdot 10^{-4}$,
  which covers
  a half of the unconstrained
  parameter region
  that explains the discrepancy in muon anomalous magnetic moment.
  The possibilities to improve the significance of
  the detection are also discussed. 
 \end{abstract}

\maketitle


\section{Introduction}
The experimental confirmation of the standard model (SM)
of particle physics was completed by the discovery
of the Higgs boson at the Large Hadron Collider (LHC)
in 2012~\cite{Aad:2012tfa,Chatrchyan:2012xdj}.
Although the SM successfully describes most of phenomena
in nature below the electroweak scale,
well-established observations such as
non zero masses of neutrinos~\cite{Fukuda:1998mi,Ahmad:2001an},
the existence of dark matter (DM)~\cite{Bertone:2004pz}
and dark energy~\cite{Perlmutter:1998np,Riess:1998cb},
and the matter-antimatter asymmetry in the Universe,
strongly require extensions of the SM.
Despite continuous and intense effort to search for new physics
at the high-energy frontier currently pushed by the LHC Run II,
any clear sign of it has not been obtained yet.
Therefore,
many attempts to discover a faint hint of new physics 
have been made and are also newly planed in the low-energy region.
In fact, extensions of the SM with a new boson possessing
a mass around the MeV scale and only feebly interacting with
our visible sector have been recently discussed in much literature,
which is motivated by phenomenological observations.
Those include
the discrepancy between the SM predictions and
the experimental measurements of 
the muon anomalous magnetic moment~\cite{Gninenko:2001hx,Baek:2001kca},
inconsistency in the measurement of the $e^{+}e^{-}$ pair
produced in the transition between an excited state of $~^{8}$Be
and its ground state~\cite{%
Krasznahorkay:2015iga,Feng:2016jff,Gu:2016ege,Feng:2016ysn,
Kitahara:2016zyb,Seto:2016pks}, 
the tension between the sterile neutrino suggested by
the short-baseline neutrino experiments and
cosmological observations~\cite{%
Hannestad:2013ana,Dasgupta:2013zpn,Tabrizi:2015bba},
the deficit of high-energy cosmic neutrino events,
which is reported by the IceCube experiment~\cite{%
Ioka:2014kca,Ng:2014pca,Ibe:2014pja,Blum:2014ewa,Araki:2014ona,Kamada:2015era,DiFranzo:2015qea,Araki:2015mya,Shoemaker:2015qul},
and the disagreement in the measurement of the proton radius~\cite{%
Barger:2010aj,TuckerSmith:2010ra,Karshenboim:2014tka,Carlson:2015poa}.
%
It is also known that
a light force carrier that intermediates between the DM
promotes the annihilation of the DM in the early Universe 
and helps to reproduce the correct relic
density~\cite{ArkaniHamed:2008qn,Feng:2010zp}.
As a theoretical framework of such a light boson, 
the gauged $U(1)^{}_{\lmlt}$ model~\cite{Foot:1990mn,He:1990pn,Foot:1994vd}
has particularly gained a lot of attention,
because the model is free from gauge anomaly 
without any extension of particle content.
Moreover, recent studies reveal that
the gauge boson with an MeV-scale mass,
which resolves the discrepancy in muon anomalous magnetic moment,
can simultaneously explain 
either the deficit in the high-energy cosmic neutrino
spectrum~\cite{Araki:2014ona,Kamada:2015era,DiFranzo:2015qea,Araki:2015mya}
or
the problem of the relic abundance of DM
in the scenario with a light weakly interacting massive
particle~\cite{Baek:2008nz,Baek:2015fea,Patra:2016shz,Biswas:2016yjr}.
%
The $\lmlt$ symmetry has been discussed also in the context
of the lepton-flavor nonuniversality in $B$
decays~\cite{Altmannshofer:2014cfa,Crivellin:2015mga,Altmannshofer:2016jzy},
lepton-flavor-violating decay of
the Higgs boson~\cite{Heeck:2014qea,Crivellin:2015mga},
the flavor structure of neutrino mass
matrix~\cite{Choubey:2004hn,Ota:2006xr,Heeck:2010pg,Heeck:2011wj,
Biswas:2016yan,Biswas:2016yjr},
and
dark matter phenomenology~\cite{Baek:2015fea,Altmannshofer:2016jzy,Biswas:2016yan,Biswas:2016yjr}.\footnote{
For phenomenology of light extra $U(1)$ gauge bosons in general,
see e.g.,
Refs.~\cite{Fayet:1990wx,Fayet:2006sp,Fayet:2007ua,Fayet:2008cn,Pospelov:2008zw,Davoudiasl:2014kua,Petraki:2014uza,Foot:2014uba,Kile:2014jea,Farzan:2015hkd, Lee:2016ief,Altmannshofer:2016brv,Farzan:2016wym,Ko:2016uft,Fayet:2016nyc}
}

In this paper, we propose a test of the models with
the $U(1)^{}_{\lmlt}$ gauge symmetry which is spontaneously
broken below the electroweak scale,
by searching for
the process $e^+e^-~\rightarrow~ \gamma+{missing~energy}
$
at the upcoming Belle-II experiment.
Among the decay channels of $Z'$, we focus on
$Z' \rightarrow \nu \bar{\nu}$ as the signal channel,
because the signal is imitated only by the weak interaction processes
and does not suffer from electromagnetic background,
as long as the final state particles are not missed at
the detector.
Since the cross section of the signal event is inversely proportional
to the square of the center-of-mass energy, the colliders with
a low energy are expected to be suitable for this type of search.
The high luminosity of the Belle-II experiment also conduces
to a good sensitivity to a feeble interaction.
The sensitivity of the B-factories to a new light gauge boson,
which is often called the dark photon,\footnote{%
We refer to
an extra $U(1)$ gauge boson as dark photon with a mass below the electroweak scale,
which couples to the SM particle content only through
the kinetic mixing with photon.
}
has been studied in the literature~\cite{Aubert:2008as,Essig:2009nc,Essig:2013vha,Lees:2017lec}
with the assumption that the kinetic mixing between the electromagnetic
$U(1)_{\rm em}$ and the $U(1)$ for the dark photon is
given as a constant parameter.
In contrast, it is not introduced as a free parameter in the minimal $U(1)_{\lmlt}$ model dealt in this paper.
In the minimal $U(1)_{\lmlt}$ model, only two parameters are newly
introduced,
which are the new gauge coupling constant
and the mass of the $Z'$.
As a consequence, the kinetic mixing arises radiatively,
and hence 
it is not a constant but depends on
the new gauge coupling and the momentum carried by the photon and the $Z'$.
We see in the next section that 
the signal event in the minimal $U(1)^{}_{\lmlt}$ model
is enhanced with the light $M_{Z'}$ with which
the discrepancy in muon anomalous magnetic moment
and the deficit in the cosmic neutrino spectrum
can be simultaneously explained.
There is no such enhancement mechanism in the models
with a constant kinetic mixing. 

This paper is organized as follows.
In the next section, we illustrate the minimal $U(1)^{}_{\lmlt}$
model and list the constraints to the model parameters
from various experiments.
The motivations for the leptonic force mediated with
such a new light gauge boson are discussed
in Sec.~\ref{Sec:Motivations-to-Zprime}.
In Sec.~\ref{Sec:Signal-and-BG},
we estimate the numbers of the signal and the background events
and 
study the feasibility of detecting the $Z'$ at the Belle-II experiment
quantitatively.
The ways to improve the sensitively
are also discussed in Sec.~\ref{Sec:Signal-Significance}.
Finally, we mention 
another type of background in Sec.~\ref{Sec:AnotherBG}
and draw conclusions in Sec.~\ref{Sec:Conclusions}.

\section{The minimal $\lmlt$ model}

Here we describe our model and list the experimental constraints
to the relevant parameter space.

\subsection{Lagrangian}
We extend the SM with a new $U(1)$ gauge symmetry associated with the
muon number minus the tau number, i.e., $U(1)_{\lmlt}^{}$, which leads
to the following new leptonic gauge interactions:
\begin{eqnarray}
\mathscr{L}_{\text{int}}=
g_{Z'}^{}Q_{\alpha\beta}(
  \overline{\ell_\alpha^{}}\gamma^\rho \ell_\beta^{} 
+ \overline{\nu_\alpha^{}}\gamma^\rho P_L \nu_\beta^{} 
)Z_\rho^\prime ,
\label{eq:Lint}
\end{eqnarray}
where $Z^\prime$ is the $U(1)_{\lmlt}^{}$ gauge boson,
$\ell_{\alpha}$ and $\nu_{\alpha}$ are charged leptons and neutrinos
with flavor $\alpha =\{e, \mu,\tau \}$, 
$g_{Z^\prime}^{}$ is the gauge coupling constant of $U(1)_{\lmlt}$, 
and the diagonal matrix $Q_{\alpha \beta} = \text{diag}(0,1,-1)$
gives the $U(1)_{\lmlt}^{}$ charges.
We assume that the $U(1)_{\lmlt}^{}$ symmetry is spontaneously broken
below the electroweak scale and the $Z'$ acquires the mass, 
\begin{eqnarray}
 {\mathscr L}_{\rm mass} = \frac{1}{2} M_{Z^\prime}^{2}
  Z^{\prime\rho}Z^{\prime}_\rho .
\end{eqnarray}
We do not introduce the kinetic mixing term
between the $U(1)_{\lmlt}^{}$ and the electromagnetic $U(1)_{\text{em}}$ gauge bosons,
which are described as~\cite{Holdom:1985ag,Foot:1991kb,Babu:1997st}
\begin{align}
 \mathscr{L}_{\rm mix} =
 -\frac{\varepsilon}{2} 
 F_{\rho \sigma} F'^{\rho \sigma},
\label{eq:L-mix}
\end{align}
where $F_{\rho \sigma}$ and $F'_{\rho \sigma}$ are the field strength
of photon and that of $Z'$,
i.e., we set $\varepsilon=0$ at the tree level.
The kinetic mixing term can be forbidden by the
introduction of the symmetry under the exchange of
$\mu$ and $\tau$,
which is held by the gauge interaction part of
the quantum electrodynamics and is softly broken
at the lepton mass terms
\cite{Foot:1994vd,Ibe:2016dir}.
In short,
our effective theory that is valid below the scale of
the $U(1)_{\lmlt}^{}$ breaking contains only two additional
parameters, which are $g_{Z^\prime}$ and $M_{Z^\prime}$.
We call this framework the
{\it the minimal $U(1)_{\lmlt}^{}$ model}.

\subsection{Experimental constraints}
\label{Sec:constraints}

\begin{figure}[t]
\includegraphics[width=\columnwidth]{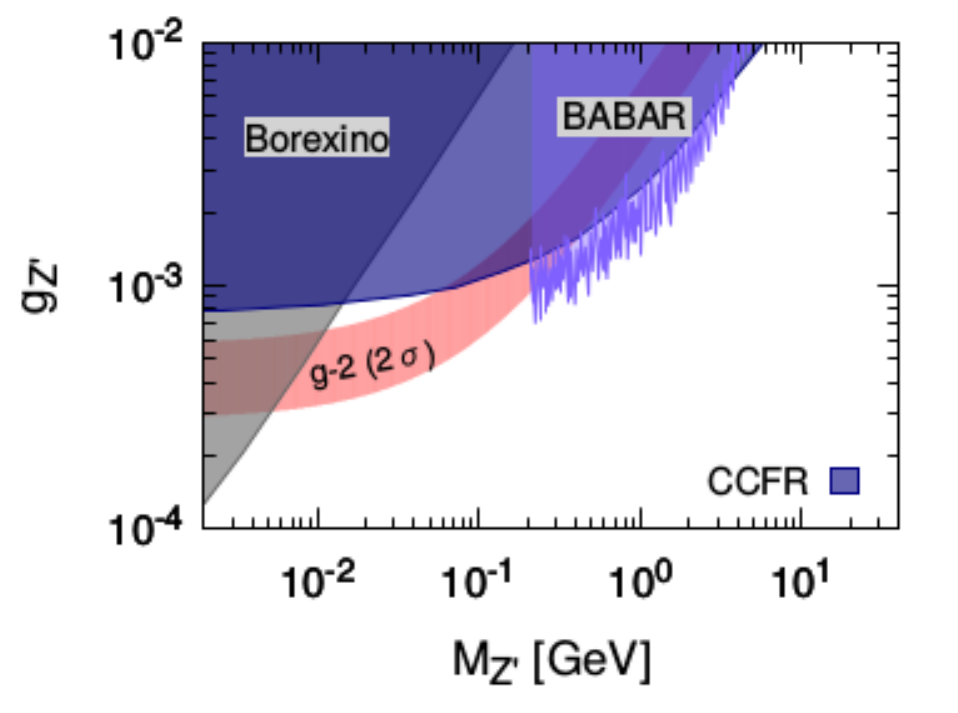}
 \caption{Summary of the parameter space of the minimal $\lmlt$ model.
 The regions shaded in blue-gray are excluded by the
 (i) neutrino-trident-production process 
 [Columbia-Chicago-Fermilab-Rochester (CCFR) experiment],
 (ii) neutrino-electron elastic scattering (Borexino detector),
 and
 (iii) muonic $Z'$ search at the collider ({\it BABAR}).
 With the parameters on the red band labeled with ``$g-2$,''
 the extra contribution from the one-loop diagram mediated by $Z'$
 resolves the discrepancy between
 the SM prediction and the experimental measurements of
 muon anomalous magnetic moment
 within $2\sigma$.
 }
\label{fig:parameter_region}
\end{figure}

As mentioned in the introduction,
our focus lies on the phenomenology of
the gauge boson $Z^\prime$ with a mass around MeV--GeV.
Such a light boson interacting with charged leptons, however,
is severely constrained by various
experiments.
In this subsection, we list the constraints relevant to
the parameter space we are interested in, which are
(i) the neutrino-trident-production process,
(ii) neutrino-electron elastic scattering,
(iii) muonic $Z'$ search in
$e^{+} e^{-} \rightarrow Z' \mu^{+} \mu^{-} \rightarrow 2 \mu^{+} 2
\mu^{-}$ at the {\it BABAR} collider,
and
(iv) $Z'$ search in meson decays.
The parameter regions excluded by those experimental results are
summarized in Fig.~\ref{fig:parameter_region}.
More discussions on the constraints can be found in
Refs.~\cite{Laha:2013xua,Harigaya:2013twa,Kamada:2015era,Araki:2015mya}
and references therein.\footnote{%
The $\lmlt$ interaction with $g_{Z'} \gtrsim 10^{-5}$
significantly decreases the diffusion rate
of neutrinos from supernova.
To circumvent the constraint from supernova cooling,
the introduction of an invisible particle that promotes
the cooling process is required~\cite{Kamada:2015era}.
}
%

The neutrino-trident-production process,
$\nu_\mu N \rightarrow \nu_\mu N \mu^+ \mu^-$
where $N$ represents a target nucleus,
is a good probe into the light $Z'$,
as pointed out in Ref.~\cite{Altmannshofer:2014pba}.
Since the cross section measured at the fixed-target neutrino
experiments~\cite{Geiregat:1990gz,Mishra:1991bv} was
found to be consistent with the SM prediction,
the contribution of the $Z'$ must be suppressed
so as to agree with the condition
\begin{eqnarray}
\frac{\sigma^{\rm CCFR}}{\sigma^{\rm SM}}=0.82 \pm 0.28 .
\label{eq:ccfr}
\end{eqnarray}
In Fig.~\ref{fig:parameter_region},
we refer to the $95\%$ C.L. limit based on the result
of the CCFR
experiment~\cite{Mishra:1991bv}.
Prospects of measuring the neutrino-trident-production process
at modern neutrino beam experiments were recently discussed
in Ref.~\cite{Magill:2016hgc} in the SM, 
and in Refs.~\cite{Kaneta:2016uyt, Ge:2017poy} in a context of $U(1)_{\lmlt}^{}$
models with the kinetic mixing at the tree level.

\begin{figure}[t]
\begin{center}
\includegraphics[clip,width=5.0cm]{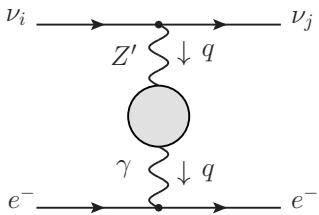}
\end{center}
 \caption{Diagram of the neutrino-electron scattering process.
 The one-loop $\gamma$-$Z'$ mixing $\varepsilon_{\nu e}$, which is expressed with a shaded blob,
 is given in Eq.~\eqref{eq:epsilon-nue}.}
\label{fig:Zp-gamma_mixing}
\end{figure}

The authors of Ref.~\cite{Harnik:2012ni} indicated that 
the precision measurement of the neutrino-electron elastic
scattering can place a stringent bound on the leptonic force
mediated by a light boson. 
Although the $Z'$ in the minimal $U(1)_{\lmlt}^{}$ model
does not couple to electrons at the tree level,
the coupling appears through the kinetic mixing
induced at the one-loop level, which
is calculated to be
 \begin{align}
  \Pi(q^{2})
  \equiv&
  \begin{minipage}{4cm}
   \unitlength=1cm
   \begin{picture}(3.5,1.5)
    \put(0,0){\includegraphics[width=3.5cm]{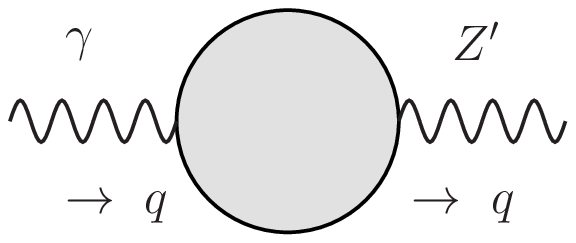}}
   \end{picture}
  \end{minipage}
 \nonumber
 \\
 =&
 \begin{minipage}{7cm}
  \unitlength=1cm
  \begin{picture}(7,1.5)
   \put(0,0.05){\includegraphics[width=7cm]{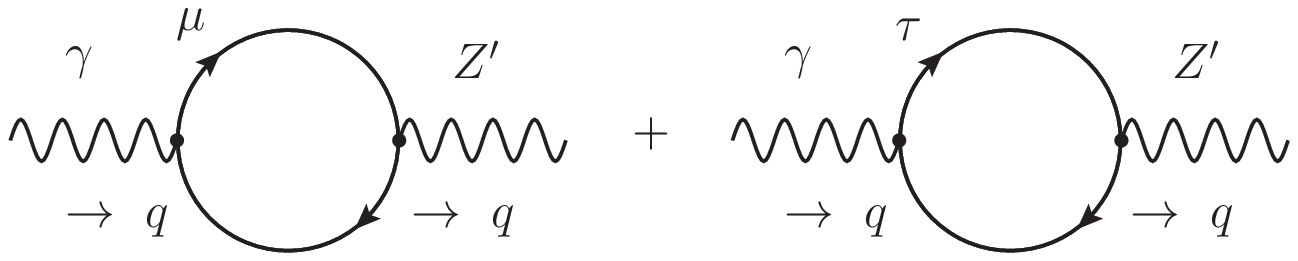}}
  \end{picture}
 \end{minipage}
 \nonumber
 \\
 =&
 \frac{8 eg_{Z^\prime}^{}}{(4\pi)^2}
 \int^1_0 x (1-x) {\rm ln}\frac{m_\tau^{2} - x(1-x)q^2}
 {m_\mu^{2} - x(1-x)q^2} ~dx,
 \label{eq:g-Zp_mixing}
 \end{align}
 where
  $e$ is the electromagnetic charge,
 $m_{\ell}$ is the mass of the charged lepton $\ell$,
 and $q$ is the momentum
 carried by $\gamma$ and $Z'$.
The kinetic mixing parameter $\varepsilon$ in Eq.~\eqref{eq:L-mix}
is given as $\varepsilon = \Pi(q^{2})$.\footnote{%
In the case where the kinetic mixing term Eq.~\eqref{eq:L-mix} exists
at the tree level, the kinetic mixing parameter $\varepsilon$
is understood as
$\varepsilon = \varepsilon_{\text{tree}}+\Pi(q^2)$~\cite{Kaneta:2016uyt}.}
With the mixing, the $Z'$ comes to contribute to the scattering
process illustrated in Fig.~\ref{fig:Zp-gamma_mixing}.
The most stringent constraint on the extra contribution
to the $\nu$-$e$ elastic scattering process is provided 
from the measurement of ${}^{7}$Be solar neutrinos 
at the Borexino detector~\cite{Bellini:2011rx}.
Since the momentum transfer $q$ in the solar neutrino
scattering process is much smaller than muon mass,
the kinetic mixing parameter $\varepsilon_{\nu e}$
relevant to this scattering process is approximately
given as
\begin{align}
 \varepsilon_{\nu e}
 = \Pi(0) =
 \frac{8}{3}
 \frac{e g_{Z'}}{(4\pi)^{2}}
 \ln \frac{m_{\tau}}{m_{\mu}}.
 \label{eq:epsilon-nue}
\end{align}
In Fig.~\ref{fig:parameter_region},
we show the bound from the Borexino experiment,
which is converted from the bound
to a gauged $U(1)_{B-L}$ model~\cite{Harnik:2012ni}.\footnote{%
The constraints to $\varepsilon_{\nu e}$ are also discussed
in Refs.~\cite{Agarwalla:2012wf,Bilmis:2015lja}.}
As we see in the next section,
the kinetic mixing parameter $\varepsilon_{\text{Belle}}$
that appears in the cross section 
of our signal process $e^{+} e^{-} \rightarrow \gamma Z'$
at the Belle-II experiment is given as 
\begin{align}
 \varepsilon_{\text{Belle}}
 = \Pi(M_{Z'}^{2}),
 \label{eq:epsilon-Belle}
\end{align}
which varies by 2 orders of magnitude according
to the mass of the $Z'$.
We emphasize that
the $q$ dependence of the kinetic mixing
makes the phenomenology of the minimal $\lmlt$ model
different from that of dark photon models
in which the kinetic mixing is given as a constant parameter.

Recently, the {\it BABAR} collaboration searched for a muonic $Z'$
in the successive processes $e^+e^- \rightarrow \mu^+\mu^- Z'$ and
$Z'\rightarrow \mu^+\mu^-$~\cite{TheBABAR:2016rlg}.
Although the signal event suffers from huge electromagnetic
backgrounds, it can be discriminated with the help of
the invariant mass distribution of the muon pairs
in the final state.
The constraint given from this process is available only
in the parameter region of $M_{Z'} > 2m_\mu$,
and we show the $90\%$ C.L.~, which
is provided in Ref.~\cite{TheBABAR:2016rlg},
in Fig.~\ref{fig:parameter_region}.

Let us briefly mention the constraints from the $Z'$ search in meson
decays.
The light $Z'$ can be produced from a muon in the final state
in decays of mesons.
The search for the $Z'$ in the charged kaon decay process
$K^+ \rightarrow \mu^+\nu_\mu Z'$
followed by
$Z' \rightarrow \nu\bar{\nu}$~\cite{Artamonov:2009sz,Artamonov:2014urb}
put the bound on the gauge coupling as $g_{Z'} \lesssim 10^{-2}$
in the relevant range of $M_{Z'}$~\cite{Ibe:2016dir},
which is much weaker than the other constraints listed above.

Finally, we make comments on light dark photon searches at 
the electron and proton beam dumps,
in which 
a pair of the charged leptons (mainly electrons) produced
in the decay of the dark photon is hunted
as the signal event.
Since 
the $Z'$ in the minimal $\lmlt$ model decays mainly to a pair of
neutrinos and the decay branching ratio to an electron pair is
negligibly small,
the constraints from the beam dump experiments
are not applicable to the minimal $\lmlt$ model
~\cite{Araki:2015mya}.
The fixed-target muon beam experiment planned by the authors
of Ref.~\cite{Gninenko:2014pea} will allow us to examine the
whole parameter region favored by the muon anomalous
magnetic moment in the $\lmlt$ model.

\subsection{Motivation to the light $Z^\prime$}
\label{Sec:Motivations-to-Zprime}
As is well known, there is a long-standing discrepancy between
the experimental measurement~\cite{Bennett:2006fi} and
the SM predictions~\cite{Melnikov:2003xd,Davier:2010nc,Hagiwara:2011af,Aoyama:2012wk,Kurz:2014wya}
of the magnetic moment of muons,
which is evaluated as 
\begin{eqnarray}
 \delta a_\mu = a_\mu^{\rm exp} - a_\mu^{\rm SM} =
  (28.7 \pm 8.0) \cdot 10^{-10},
  \label{eq:delta-a-current}
\end{eqnarray}
in terms of $a_\mu \equiv (g_\mu-2)/2$.
The new interaction with muons, which is introduced in Eq.~\eqref{eq:Lint},
provides an extra contribution to $a_\mu$,
which is calculated
as~\cite{Gninenko:2001hx,Baek:2001kca}\footnote{%
The introduction of the tree-level kinetic mixing 
$\varepsilon_{\text{tree}}$ changes the gauge coupling for muon 
from $g_{Z'}$ to $e \varepsilon_{\text{tree}} + g_{Z'}$.
The region excluded by the CCFR and {\it BABAR} experiments and the region favored
by muon $g-2$ shown in Fig.~\ref{fig:parameter_region}
are shifted by this change of the coupling.
The cross section of the neutrino-electron scattering process
at Borexino is multiplied by
$|\varepsilon_{\text{tree}} + \varepsilon_{\nu e}|^{2}
/|\varepsilon_{\nu e}|^{2}$.
For more discussion on the parameter region of the $\lmlt$ model
with the tree-level kinetic mixing, see Ref.~\cite{Kaneta:2016uyt}.
}
\begin{eqnarray}
a_\mu^{Z^\prime} = 
\frac{g_{Z^\prime}^2}{8\pi^2}
\int^1_0 \frac{2m_\mu^2 x^2(1-x)}{x^2m_\mu^2 + (1-x)M_{Z^\prime}^2}dx .
\end{eqnarray}
The parameter region on which
the $Z'$ contribution resolves the discrepancy
in the muon anomalous magnetic moment
at $2\sigma$ is indicated with the red band (labeled with $g-2$)
in Fig.~\ref{fig:parameter_region}.
After the constraints listed in the previous subsection
are taken into consideration,
a narrow window of the parameter region 
 \begin{align}
  M_{Z^\prime} &\simeq [5\cdot 10^{-3} , 2\cdot 10^{-1}]~{\rm GeV}
  \nonumber\\
  g_{Z^\prime} &\simeq [3\cdot10^{-4} , 1\cdot 10^{-3}]
\label{eq:para_space}
 \end{align}
which is favored by the muon $g-2$,
is still allowed.

It is interesting to point out that the $Z^\prime$ lies on the parameter
region of Eq.~\eqref{eq:para_space}, resonantly enhancing the scattering
of high-energy cosmic neutrinos on the cosmic neutrino
background, and the scattering can leave characteristic absorption lines
in the cosmic neutrino spectrum observed at the Earth~\cite{Araki:2014ona}.
The IceCube experiment reported a gap in the cosmic
neutrino spectrum between $400$ TeV and
$1$ PeV~\cite{Aartsen:2014gkd},\footnote{
In the four-year IceCube data~\cite{Aartsen:2015zva}
the gap becomes narrower but still exists.}
and it was demonstrated in Ref.~\cite{Araki:2015mya}
that the IceCube gap and the discrepancy in the muon anomalous
magnetic moment can be simultaneously resolved by
the $\lmlt$ force with a set of the parameters
in the range of Eq.~\eqref{eq:para_space}.

\section{Light $Z^\prime$ search at Belle-II}
\label{Sec:Signal-and-BG}
We study the feasibility to detect the $Z'$ at
the future Belle-II experiment, which is an electron-positron
collider with the center-of-mass energy of $\sqrt s$=10.58 GeV
designed to achieve the integrated luminosity of
50 ${\rm ab}^{-1}$ by the middle of the next decade.
Although the observation of the muon anomalous magnetic moment
favors the parameter region shown in Eq.~\eqref{eq:para_space},
for the sake of completeness, we broaden
our scope of $M_{Z'}$ to $[0, 10]$ GeV, 
which is the mass range possible to be explored
at the Belle-II experiment.
%

\subsection{Signal:
  $e^+e^- \rightarrow \gamma Z^\prime$, $Z^\prime \rightarrow \nu\bar{\nu}$}
\begin{figure}[t]
\begin{center}
\includegraphics[clip,width=6.0cm]{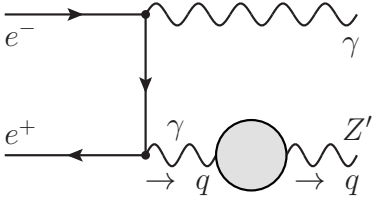}
\end{center}
\caption{
 Diagram of the signal process at the Belle experiment:
 The $Z'$ production $e^{+}e^{-}\rightarrow \gamma Z'$
 through the $\gamma$-$Z'$ mixing $\varepsilon_{\text{Belle}}$
 given in Eq.~\eqref{eq:epsilon-Belle}.
 }
\label{fig:gamma_missing_process}
\end{figure}
\begin{figure}[t]
\includegraphics[width=\columnwidth]
{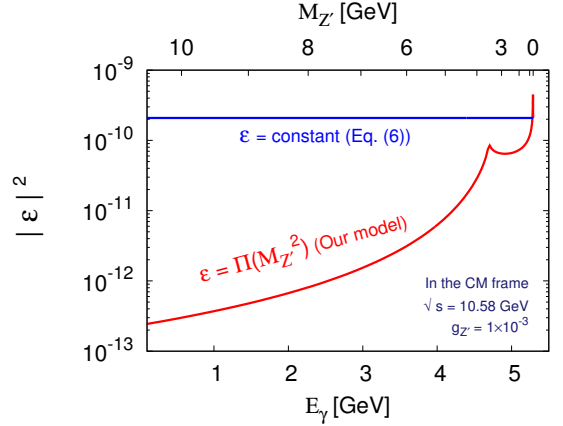}
 \caption{
 The $\gamma$-$Z'$ mixing $\Pi$ as a function of
 the photon energy $E_{\gamma}$ (red).
 The coupling $g_{Z'}$ is taken to be $1 \cdot 10^{-3}$.
 The upper horizontal axis represents $M_{Z'}$
 which is related to $E_{\gamma}$
 as Eq.~\eqref{eq:Eg-q2}.
 We also show $\varepsilon_{\nu e}$ in blue,
 which is the kinetic mixing parameter appearing
 in the neutrino-electron elastic scattering
 and is given in Eq.~\eqref{eq:epsilon-nue},
 as a comparison.
 }
\label{fig:Pi_square}
\end{figure}

With the interaction given in Eq.~\eqref{eq:Lint},
the $Z'$ is produced on its mass shell
through the diagram shown in Fig.~\ref{fig:gamma_missing_process}.\footnote{%
Note that the one-loop triangle diagrams
that prompt $e^{+} e^{-} \rightarrow \gamma^{*} \rightarrow \gamma Z'$
are canceled with each other due to the Furry's theorem.
The same type of cancellation mechanism in the radiative
$Z \rightarrow \gamma\gamma$ decay is discussed in
Ref.~\cite{Zhemchugov:2014dza} and references therein.
}
Depending on its mass, the $Z'$ subsequently decays 
not only into a pair of neutrinos 
but also into charged leptons.
In this study, we focus on the $Z'\rightarrow \nu\bar{\nu}$ decay mode,
because the process $e^{+}e^{-}\rightarrow \gamma +\not{E}$,
where $\not{E}$ denotes missing energies carried by neutrinos,
does not suffer from electromagnetic backgrounds,
if the final state particles are not missed by the detectors.
The $Z'$ is produced through the kinetic mixing,
which is given in Eq.~\eqref{eq:g-Zp_mixing}
as a function of the momentum $q$ carried by
$Z'$.
In Fig.~\ref{fig:Pi_square},
we plot the square of the kinetic mixing
$|\varepsilon_{\text{Belle}}|^{2}=|\Pi(q^{2}=M_{Z'}^{2})|^{2}$ 
as a function of $E_\gamma$,
where $E_\gamma$ denotes the energy of the final state photon
and is related to $q^2$ as
\begin{align}
E_{\gamma}  = \frac{s-q^{2}}{2\sqrt{s}}
\label{eq:Eg-q2}
\end{align}
in the center-of-mass frame.
Here 
the gauge coupling is taken to be $g_{Z'} = 1.0 \cdot 10^{-3}$.
We also show the value of $\varepsilon_{\nu e}$
given in Eq.~\eqref{eq:epsilon-nue} 
as a comparison. 
The loop-induced kinetic mixing
can change by 2 orders of magnitude
in the range of $M_{Z'}$, which Belle-II can explore.
This feature distinguishes the phenomenology of the $\lmlt$ model
from the dark photon models with a constant value of the kinetic mixing.
It is clearly recognizable in Fig.~\ref{fig:Pi_square}
that the kinetic mixing $\Pi(q^{2})$
is enhanced with a large value of $E_{\gamma}$,
which corresponds to a light $Z'$.

\begin{figure}[t]
\includegraphics[width=\columnwidth]
{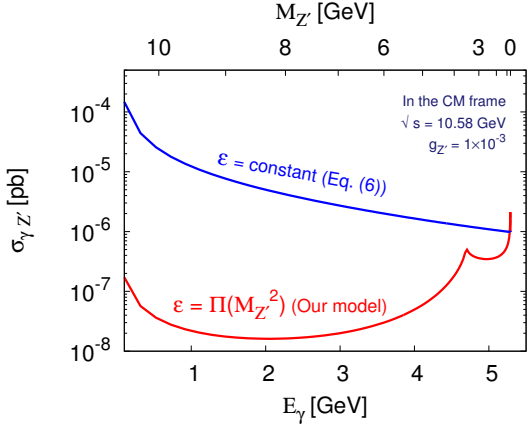}
 \caption{The cross sections of the $Z'$ production process
 $e^{+} e^{-} \rightarrow \gamma Z'$
 as a function of $E_\gamma$ (lower axis) and $M_{Z'}$ (upper axis).
 The red and the blue curves correspond
 to the minimal $U(1)_{\lmlt}^{}$ model
 and the case with the constant value $\varepsilon_{\nu e}$
 for the kinetic mixing, respectively.
 }
\label{fig:contrast}
\end{figure}

The differential cross section of the signal process
$e^+e^- \rightarrow \gamma Z'$
in the center-of-mass frame
is found to be~\cite{Boehm:2003hm,Borodatchenkova:2005ct,Essig:2009nc}
 \begin{align}
  \hspace*{-0.2cm}
 \frac{d\sigma_{\gamma Z'}}{d\cos \theta}=
 & \frac{2\pi\alpha^{2} |\Pi(M_{Z'}^{2})|^{2}}{s}
 \left[
 1-\frac{M_{Z'}^{2}}{s}
 \right]
 \frac{
 1+\cos^{2}\theta +
 \frac{4sM_{Z'}^{2}}{(s-M_{Z'}^{2})^{2}}
 }{(1+\cos\theta)(1-\cos\theta)}\ ,
 \label{diff_cs_gammaZprime}
\end{align}
where $\alpha$ is the fine structure constant and $\theta$ is
the angle between the electron beam axis and the photon momentum.
The cross section after integrating
the angle $\theta$ over the range of
the coverage of the electromagnetic calorimeters
is given as~\cite{Essig:2009nc}
\begin{align}
 \sigma_{\gamma Z'}
 =&
 \frac{2\pi\alpha^{2} |\Pi(M_{Z'}^{2})|^{2}}{s}
 \left[
 1-\frac{M_{Z'}^{2}}{s}
 \right]
 \nonumber
 \\
 \times&
 \Biggl[
 \left[1+\frac{2sM_{Z'}^{2}}{(s-M_{Z'}^{2})^{2}} \right]
 \ln
 \frac{(1+\cos\theta_{\rm max})(1-\cos\theta_{\rm min})}{(1-\cos\theta_{\rm max})(1+\cos\theta_{\rm min})}
 \nonumber
 \\
 &\hspace{0.5cm} 
 -\cos\theta_{\rm max}+\cos\theta_{\rm min}
 \Biggr],
 \label{cs_gammaZprime}
\end{align} 
where $\cos\theta_{\rm min} = 0.941$ and
$\cos\theta_{\rm max} = -0.821$ in the center-of-mass frame
of the Belle-II experiment.
In Fig.~\ref{fig:contrast}, the cross section
in the minimal $\lmlt$ model is compared to
that calculated with the kinetic mixing of
the constant value $\varepsilon_{\nu e}$ given
in Eq.~\eqref{eq:epsilon-nue}.
The cross section of the minimal $\lmlt$ models
is enhanced in the high $E_{\gamma}$ region due to the $q^2$ dependence
in $\Pi(q^{2})$, as illustrated
in Fig.~\ref{fig:Pi_square}.
The coupling $g_{Z'}$ is taken to be $10^{-3}$ in the plot,
and the cross section is scaled as $g_{Z'}^{2}$ as
seen in the analytic expression of the cross section,
Eq.~\eqref{cs_gammaZprime}.
%

Since the $Z'$ can decay not only into a pair of neutrinos
but also to charged leptons, 
the rate for the signal process
$e^{+}e^{-}\rightarrow \gamma Z'$, $Z' \rightarrow \nu\bar{\nu}$
is calculated by multiplying
the cross section $\sigma_{\gamma Z'}$ in Eq.~\eqref{cs_gammaZprime}
by the branching ratio
for the $Z' \rightarrow \nu\bar{\nu}$ process,
which is given as\footnote{%
The $Z'$ can decay also into a pair of electrons through the kinetic
mixing. However, the branching ratio is negligibly small.}
\begin{align}
{\rm Br}(Z' \rightarrow \nu\bar{\nu})
 &=
 \begin{cases}
  1, & \hspace{-0.2cm}(M_{Z'} < 2m_\mu),
  \vspace{0.2cm}
  \\
  \cfrac{\Gamma(Z' \rightarrow \nu\bar{\nu})}
  {\displaystyle
  \sum_{f=\nu, \mu}
  \hspace{-0.1cm}
  \Gamma(Z' \rightarrow f\bar{f})},
  & \hspace{-0.2cm}(2m_\mu < M_{Z'} < 2m_\tau),
  \vspace{0.2cm}
  \\
  \cfrac{\Gamma(Z' \rightarrow \nu\bar{\nu})}
  {\displaystyle
  \sum_{f=\nu,\mu,\tau}
  \hspace{-0.25cm}
  \Gamma(Z' \rightarrow f\bar{f})},
  & \hspace{-0.2cm}(2m_\tau < M_{Z'}).
 \end{cases}
\end{align}
The decay rates are calculated to be
\begin{align}
 \Gamma(Z' \rightarrow \nu\bar{\nu}) 
 & = \frac{g_{Z'}^{2}}{12\pi}M_{Z'},
 \\
 \Gamma(Z' \rightarrow \ell^{+}\ell^{-})
 & =  \frac{g_{Z'}^{2}}{12\pi}M_{Z'}
 \left[
 1+\frac{2m_{\ell}^{2}}{M_{Z'}^{2}}
 \right]
 \sqrt{1-\frac{4m_{\ell}^{2}}{M_{Z'}^{2}}},
\end{align}
where $\ell = \{\mu, \tau\}$.

\subsection{SM background}
\begin{figure}[t]
  \includegraphics[width=6.0cm]{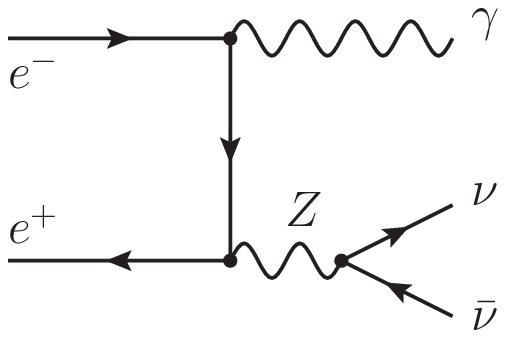}
  \\
  \includegraphics[width=6.0cm]{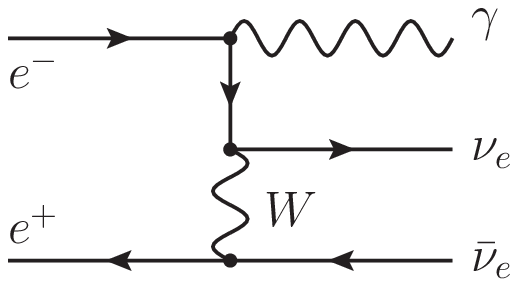}
  \\
  \includegraphics[width=6.0cm]{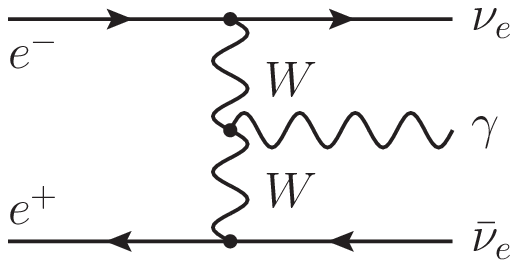}
  \caption{
  Diagrams of the SM background.
  The $W$ boson diagrams produce only electron neutrinos,
  while the $Z$ boson diagram does 
  all the flavors of neutrinos.
  }
  \label{fig:gamma_missing_SM}
 \end{figure}

The signal process $e^{+}e^{-}\rightarrow \gamma + \not{E}$
is also replicated with the SM processes mediated
by an off-shell weak boson, which are shown
in Fig.~\ref{fig:gamma_missing_SM}.
They provide the inevitable background event.\footnote{%
We discuss the background events caused by
failing to detect the final state particles  
in Sec.~\ref{Sec:AnotherBG}.
We call the $e^{+} e^{-} \rightarrow \gamma \nu \bar{\nu}$ process
mediated by the weak gauge bosons (shown in Fig.~\ref{fig:gamma_missing_SM})
the SM BG.
}
The diagram shown in the bottom
of Fig.~\ref{fig:gamma_missing_SM},
in which the final state
photon is emitted from the $WW\gamma$ vertex,
can be safely eliminated from our evaluation of the background,
because the diagram is suppressed by an additional $W$ boson propagator
in comparison with the other diagrams.
The background processes with muon and tau neutrinos in the final
state are led only from 
the diagram mediated by a $Z$ boson
(top of Fig.~\ref{fig:gamma_missing_SM}).
On the other hand, 
all the diagrams contribute to the process
with a pair of electron neutrinos.
The differential cross section
of the SM background (BG) is given as
\begin{align}
\frac{d\sigma_{\rm SM}}{dE_{\gamma}}
 =&
 \frac{\alpha G_{F}^{2}}{3\pi^{2}} 
 (g_{L}^{2}+g_{R}^{2}) E_{\gamma}
 \left[
 1-\frac{2E_{\gamma}}{\sqrt{s}}
 \right]
 \nonumber
 \\
 \times&
 \Biggl[
 \left[
 1
 -
 \frac{\sqrt{s}}{E_{\gamma}}
 +
 \frac{s}{2E_{\gamma}^{2}}
 \right]
 \ln
 \frac{(1+\cos\theta_{\rm max})(1-\cos\theta_{\rm min})}
 {(1-\cos\theta_{\rm max})(1+\cos\theta_{\rm min})}
 \nonumber
 \\
 &
 -
 \cos\theta_{\rm max}+\cos\theta_{\rm min}
 \Biggr]
\label{diff_cs_SM}
\end{align}
in the center-of-mass frame, where the couplings
are defined as
\begin{align}
g_{L}& =
\begin{cases}
-\frac{1}{2}+\sin^{2}\theta_{W} & ({\rm for\ }\nu_{\mu},\nu_{\tau})\\
-\frac{1}{2}+\sin^{2}\theta_{W}+1 & ({\rm for\ }\nu_{e})
  \end{cases}
\notag \\
g_{R} & = \sin^{2}\theta_{W},
\end{align}
and $\theta_W$ is the Weinberg angle.
We have checked that Eq.~\eqref{diff_cs_SM}
is consistent with the result reported in Ref.~\cite{Ma:1978zm}.
%

\subsection{Signal significance}
\label{Sec:Signal-Significance}
\begin{figure}[t]
\includegraphics[width=\columnwidth]{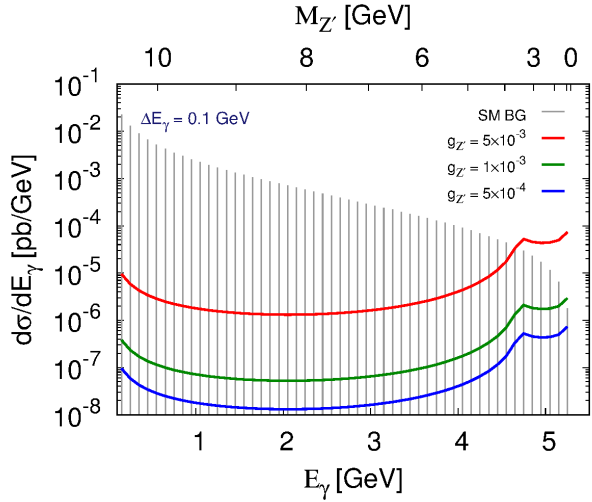}
\caption{
 The differential cross sections as functions of $E_\gamma$.
 The solid curves show the cross sections
 of the signal process with different values
 of the couplings
 $g_{Z'} = \{5\cdot 10^{-4}\text{ (blue)}, 1\cdot 10^{-3}\text{ (green)}, 5\cdot
 10^{-3}\text{ (red)}\}$.
 The cross sections are calculated with the value of $M_{Z'}$
 indicated with the top horizontal axis.
 The $E_{\gamma}$ dependence of the SM background cross section
 is shown as the region shaded with gray vertical stripes.
 }
\label{fig:diff-cross-sec}
\end{figure}
\begin{figure}[t]
\includegraphics[width=\columnwidth]{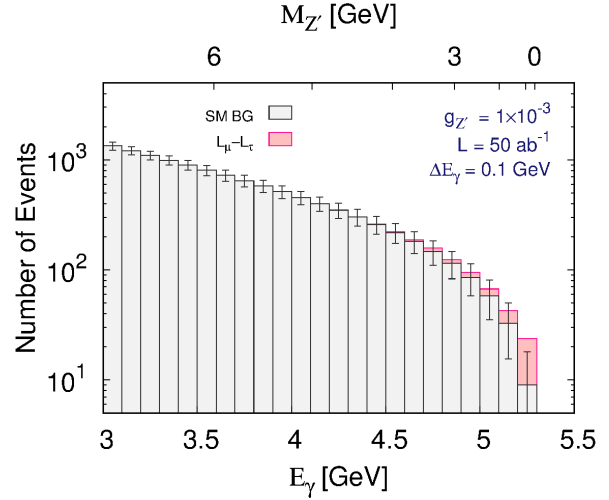}
 \caption{
 $E_{\gamma}$ distribution of the event numbers.
 The red histogram shows the number of
 the signal events $N_{\text{sig}}$ calculated
 with the coupling
 $g_{Z'} = 10^{-3}$ and the mass $M_{Z'}$ indicated
 with the upper horizontal axis,
 while
 the gray shows the SM background events $N_{\text{BG}}$.
 The integrated luminosity is assumed to be $50$ ab$^{-1}$.
 The error bars indicate the range of the $3\sigma$ statistical
 error.
 }
\label{fig:number-of-events}
\end{figure}

The $E_\gamma$ dependence of the differential
cross sections of the signal process
$e^{+} e^{-} \rightarrow \gamma Z' \rightarrow \gamma \nu \bar{\nu}$
is compared to that of the SM background
in Fig.~\ref{fig:diff-cross-sec}.
The cross section Eq.~\eqref{cs_gammaZprime} of the signal process
is enhanced in the high $E_\gamma$ region, 
due to the $q^2$ dependence of the $\Pi$ function
(cf. Fig.~\ref{fig:Pi_square}),
while the SM background is suppressed.
Figure \ref{fig:diff-cross-sec} shows that
the signal with the coupling $g_{Z'} \gsim 10^{-3}$ 
becomes larger than the SM background 
around the high $E_\gamma$ end.
We emphasize again that
the $E_{\gamma}$ dependence (equivalent to the $M_{Z'}$ dependence)
of the signal is a characteristic feature of the minimal $\lmlt$ model
and is different from the dark photon models with a constant
kinetic mixing.
The signal and the background are compared also
in their numbers of event in Fig.~\ref{fig:number-of-events},
where the red histogram shows the $M_{Z'}$ dependence
($E_{\gamma}$ dependence) of
the signal event $N_{\text{sig}}$
and the gray shows the $E_{\gamma}$ distribution of
the SM background event $N_{\text{BG}}$,
respectively.
The integrated luminosity $\mathcal{L}$ is assumed to be
$50$ ab$^{-1}$.
The detector resolution to the photon energy, 
which is understood also as the width of each energy bin,
is taken to be
\begin{equation}
 \Delta E_\gamma =0.1~{\rm GeV}
\end{equation}
\cite{Abe:2010gxa}.
Here, we assume that the energy resolution in the center-of-mass
frame is the same as the one in the laboratory frame.
The error bars indicate the range of the $3 \sigma$ statistical
error estimated with the square root of the number of the background
event, $\sqrt{N_{\mathrm{BG}}}$.
The number of the signal event with the coupling
$g_{Z'} = 10^{-3}$ exceeds the SM background
more than $3\sigma$ in the highest energy bin,
which corresponds to the $Z'$ with $M_{Z'} \lesssim 1$ GeV.
From this result, we can expect that the Belle-II experiment will
be sensitive to the light $Z'$ ($M_{Z'} \lesssim 1$ GeV)
with the coupling $g_{Z'} \gtrsim 10^{-3}$.

\begin{figure}[t]
\includegraphics[width=\columnwidth]{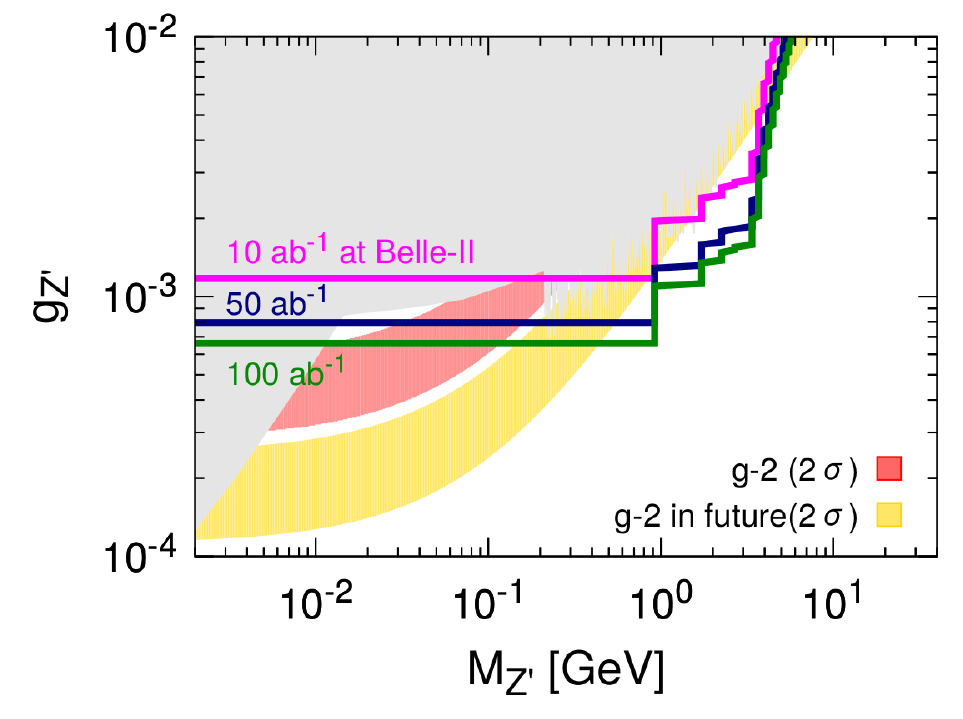}
\caption{
 Parameter regions with the signal significance $\mathcal{S}$
 larger than $3$.
 The integrated luminosity is taken to be $10$ (magenta),
 $50$ (blue) and $100$ (green) ab$^{-1}$.
 The experimental constraints
 summarized in Fig.~\ref{fig:parameter_region}
 are shown in gray.
 The red and yellow bands indicate
 the parameter regions favored
 by the current [Eq.~\eqref{eq:delta-a-current}]
 and the future [Eq.~\eqref{eq:delta-a-future}]
 muon $g-2$ measurements.
}
\label{fig:3sigma-region1}
\end{figure}

\begin{figure}[t]
\includegraphics[width=\columnwidth]{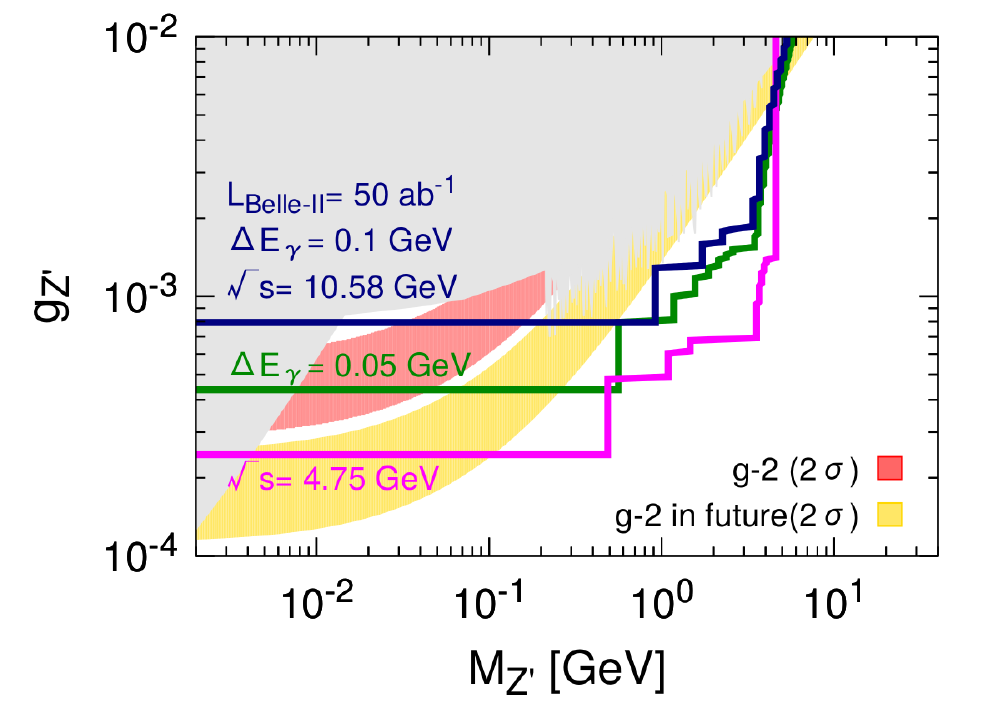}
\caption{%
 The same plot as Fig.~\ref{fig:3sigma-region1}
 with $\Delta E_\gamma = 0.05$ GeV (green) and
 $\sqrt{s} = 4.75$ GeV (magenta). 
 These correspond to the setup given in  
 Eq.~\eqref{parameter_setup}. 
 }
 \label{fig:3sigma-region2}
\end{figure}

In order to illustrate the parameter region
on which the Belle-II experiment can detect the $Z'$
through the signal process $e^{+} e^{-} \rightarrow \gamma + \not{E}$,
we define the signal significance $\mathcal{S}$ as 
\begin{align}
\mathcal{S} \equiv \frac{N_{\mathrm{sig}}(g_{Z'}, M_{Z'})}{\sqrt{N_{\mathrm{BG}}}}
\end{align}
and adopt $\mathcal{S} > 3$ for the criterion of
the signal detection,
i.e.,
we expect that 
Belle-II will be sensitive to 
the parameter sets $(g_{Z'}, M_{Z'})$ that satisfy the criterion.
In Fig.~\ref{fig:3sigma-region1},
we draw the boundaries of the parameter regions
that will be examined by the Belle-II experiment
with three different integrated luminosities
$\mathcal{L} = \{10, 50, 100\}$ ab$^{-1}$.
The signal significance $\mathcal{S}$ exceeds 3
on the regions of the upper side of each curve.
The regions shaded in gray are excluded by the
experimental constraints listed in Sec.~\ref{Sec:constraints}
(CCFR, Borexino, and {\it BABAR}).
The red band indicates the parameters
favored by muon $g-2$ within $2\sigma$.
We also show in yellow the parameter region favored by
the discrepancy in $a_{\mu}$ with the value of 
\begin{equation}
 \delta a_\mu = (4.8 \pm 1.6) \cdot 10^{-10},
  \label{eq:delta-a-future}
\end{equation}
where we expect 
that the error in $a_{\mu}$ will be reduced by a factor of 5
in future experiments~\cite{Grange:2015fou,KEKE34} 
and assume that the discrepancy will be kept at 3 $\sigma$.
It is shown that the Belle-II experiment with
the full integrated luminosity is expected to be sensitive
to the parameter region with $M_{Z'} \lesssim 1$ GeV and
$g_{Z'} \gsim 8 \times 10^{-4}$.
The sensitivity becomes improved with a higher luminosity
such as $\mathcal{L} = 100$ ab$^{-1}$,
because the significance is proportional to $\sqrt{\mathcal{L}}$.
We examine the possible improvements in the sensitivity
with the change of the energy resolution $\Delta E_{\gamma}$
and the center-of-mass energy $\sqrt{s}$.
We compare the sensitivity reaches estimated
with the following two sets of parameters,
 \begin{align}
  (\Delta E_\gamma, \sqrt{s})
  =
  \begin{cases}
   (0.05 \text{ [GeV]}, 10.58 \text{ [GeV]}), \quad \text{(green)},
   \\
   (0.1 \text{ [GeV]}, 4.75 \text{ [GeV]}), \quad \text{(magenta)}, 
  \end{cases} 
  \label{parameter_setup}
 \end{align} 
in Fig.~\ref{fig:3sigma-region2}.
The event number of the signal process is inversely proportional to $s$
[cf., Eq.~\eqref{cs_gammaZprime}],
\begin{equation}
N_{\mathrm{sig}} \propto 1/s,
\end{equation} 
and hence it is enhanced with a lower value of the center-of-mass
energy,
while the SM background depends on $\sqrt{s}$ and $\Delta E_{\gamma}$
as [cf., Eq.~\eqref{diff_cs_SM}]
\begin{equation}
 N_{\mathrm{BG}} 
 \propto \sqrt{s} \Delta E_\gamma,
\end{equation}
which is reduced with a lower $\sqrt{s}$ and
a smaller $\Delta E_\gamma$.
Figure ~\ref{fig:3sigma-region2} shows that 
the region favored by the current measurement of muon $g-2$
at $2\sigma$ is completely covered by the sensitivity reach of
the Belle-II experiment
with the center-of-mass energy $\sqrt{s}=4.75$ GeV
and $\mathcal{L}=50\ {\rm ab}^{-1}$. 

  \subsection{Another possible background:
  $e^{+} e^{-} \rightarrow \gamma + undetected$ }
\label{Sec:AnotherBG}
  
We have estimated the number of the inevitable
background event mediated by the weak interaction
and discussed the signal significance
and the sensitivity reach in the previous
subsections.
However, because of the limitation in detector coverage and
detection efficiency,
it is possible that 
the electromagnetic processes $e^{+} e^{-} \rightarrow \gamma +X$ 
with undetected final states $X$ come into backgrounds.
The undetected state $X$ can be $n \gamma$ ($n \ge 1$),
$e^+e^-$, etc.
It is pointed out in Refs.~\cite{Aubert:2008as,Essig:2013vha,Lees:2017lec}
that the process with $X=\gamma$ can be a serious background
in the detection of light $Z'$, 
because the signal photon shows the same kinematics
as the photons in the background event
up to the order of $M_{Z'}^{2}/s$.
Since the undetection rate of photon at the Belle-II experiment
strongly depends on its detector properties,
an estimation of the rate requires a dedicated detector simulation.
In this study, we expect that the detection efficiency
for photons with energies of $\sqrt{s}/2$ is high enough 
to reconstruct perfectly 
the back-to-back two photon events,
$e^{+} e^{-} \rightarrow \gamma \gamma$.
We carry out a numerical simulation,
which is specialized to the experimental setup of Belle II,
to optimize the kinematical cuts so as to
maximize the signal significance,
in a forthcoming paper.

\section{Conclusions and discussion}
\label{Sec:Conclusions}

We have discussed the sensitivity of the Belle-II experiment
to the light gauge boson $Z'$
in the minimal gauged $U(1)^{}_{\lmlt}$ model.
With the new gauge interaction,
the $Z'$ is produced through the kinetic mixing between
photon and $Z'$, which is absent at the tree level but is
induced radiatively.
We have focused on 
$e^{+} e^{-} \rightarrow \gamma Z' \rightarrow \gamma \nu \bar{\nu}$
as the signal process,
because it does not suffer from
a huge number of electromagnetic background events,
as long as the undetection rate of high-energy photons
is sufficiently suppressed.
Differing from the search for a dark photon with a constant kinetic mixing
(e.g., Ref.~\cite{Essig:2013vha}),
the signal event in the $U(1)^{}_{\lmlt}$ model is strongly enhanced
with a small $M_{Z'}$, with which the inconsistency in the muon anomalous
magnetic moment and the gap in the IceCube spectrum can be
simultaneously explained~\cite{Araki:2014ona}.
The cross section of the signal event reaches $\mathcal{O}(1)$
ab with the parameters $g_{Z'} = 10^{-3}$ and $M_{Z'} < 1$ GeV,
i.e., $\mathcal{O}(10)$ events are expected 
at the Belle-II experiment with the design luminosity.
We have estimated the number of background events
and calculated the signal significance 
to illustrate the parameter region to which
the Belle-II experiment will be sensitive.
The
SM
background events are distributed more
in the lower energy regions of the final state photon.
We have shown that the signal events with
a high-energy photon, which corresponds to a low $M_{Z'}$,
can be discriminated from the background events.
We have found that 
the Belle-II experiment with the design luminosity
can examine a part of the parameter region
that evades the current experimental constraints 
and, at the same time, is favored by the observation of
the muon anomalous magnetic moment.
We have shown that the further improvement in sensitivity
is possible with
an increase of the luminosity,
an adjustment of the center-of-mass energy,
and the upgrading of the energy resolution of the calorimeter.

As a concluding remark, we emphasize that our analysis can be easily generalized to any models with a light $Z'$ that has an invisible decay channel and loop-induced kinetic mixing. 

\begin{acknowledgments}
 The authors are particularly grateful
 to Professor Kiyoshi Hayasaka
 for the information about
 the experimental setup of Belle-II.
This work is supported by JSPS KAKENHI Grants 
No.~25105009 (J.S.) and 
No.~15K17654 (T.S.),
and
the Sasakawa Scientific Research Grant 
No.~28-210 (S.H.)
from the Japan Science Society.
T.S. thanks the visitor support program in
the Japan Particle and Nuclear Forum.
 The authors express their gratitude
 to MEXT Grant-in-Aid for Scientific Research on Innovative Areas
 {\it Unification and Development of the Neutrino Science Frontier}
 for its support.
\end{acknowledgments} 

\bibliography{./Lmu-Ltau_BelleII}
\bibliographystyle{apsrev}

\end{document}